\documentclass[%
 reprint,longbibliography,preprintnumbers,
nofootinbib,
 amsmath,amssymb,
 aps,
pre,
]{revtex4-1}
\pdfoutput=1
\usepackage[utf8]{inputenc}
\usepackage{flushend}
\usepackage{dcolumn}
\usepackage{bm}
\usepackage{balance}

\usepackage[normalem]{ulem}

\usepackage[colorlinks = true,
            linkcolor = blue,
            urlcolor  = blue,
            citecolor =green,
            anchorcolor = blue]{hyperref}
\usepackage{verbatim}
\usepackage{color,ulem}
\usepackage[english]{babel}

\usepackage[utf8]{inputenc}
\input Starburst.fd
\newcommand*\initfamily{\usefont{U}{Starburst}{xl}{n}}\initfamily

\newcommand{\beq}{\begin{eqnarray}}
\newcommand{\eeq}{\end{eqnarray}}
\usepackage{amsmath}
\usepackage{tikz}
\usetikzlibrary{decorations.pathmorphing}
\usetikzlibrary{shapes.misc}
\tikzset{cross/.style={cross out, draw=black, minimum size=8*(#1-\pgflinewidth), inner sep=0pt, outer sep=0pt},
cross/.default={1pt}}
\usetikzlibrary{patterns,math}
\begin{document}

\title{General theory of the viscosity of liquids and solids from nonaffine particle motions}

\author{\textbf{Alessio Zaccone}$^{1}$}%
 \email{alessio.zaccone@unimi.it}
 
 \vspace{1cm}
 
\affiliation{$^{1}$Department of Physics ``A. Pontremoli'', University of Milan, via Celoria 16,
20133 Milan, Italy.}

\begin{abstract}
A new microscopic formula for the viscosity of liquids and solids is derived rigorously from a first-principles (microscopically reversible) Hamiltonian for particle-bath atomistic motion. The derivation is done within the framework of nonaffine linear response theory. The new formula may lead to a valid alternative to the Green-Kubo approach to describe the viscosity of condensed matter systems from molecular simulations without having to fit long-time tails. Furthermore, it provides a direct link between the viscosity, the vibrational density of states of the system, and the zero-frequency limit of the memory kernel. Finally, it provides a microscopic solution to Maxwell's interpolation problem of viscoelasticity by naturally recovering Newton's law of viscous flow and Hooke's law of elastic solids in two opposite limits.
\end{abstract}

\maketitle
\section{Introduction}
The viscosity of a system is a measure of both its internal anharmonic
forces as well as of the random motion of its underlying microscopic constituents. As a ubiquitous transport coefficient, it determines many properties of a material \cite{Hansen}, including sound attenuation \cite{Landau_elasticity,Baggioli_2022} and nonlinear phenomena such as shock wave propagation \cite{Vitelli}.

Understanding viscosity at a microscopic level is also crucial to solving outstanding mysteries in physics, such as the so-called viscosity minimum \cite{Murillo} and 
the existence of non-dissipative fluids described by the
Euler hydrodynamics such as quark-gluon plasmas \cite{quark} and the microscopic mechanism of superfluidity in helium-4 \cite{Trachenko_2023}. It also plays a crucial role in the glass transition of supercooled liquids \cite{KSZ,Ginzburg1,Ginzburg2} and in the viscoelasticity of confined liquids \cite{PNAS2020,Riedo}.

From the point of view of applications, the viscosity of liquids controls the diffusive aggregation and gelation of colloids \cite{Joep} via the Stokes-Einstein diffusion coefficient, and, via the Peclet number, plays a crucial role in the colloidal stability of suspensions under shear flows \cite{Zaccone2009,Wu} and in the microstructure of nonequilibrium liquids under shear flow \cite{Banetta1,Banetta2}. In biology, viscosity is a key parameter to understand and explain physico-chemical conditions that are favourable for life processes \cite{Kostya_preprint}.

The viscosity $\eta$ of liquids controls kinetic energy dissipation under flow as $\dot{E}_{kin}\sim -\eta \dot{\gamma}^{2}$, where $\dot{\gamma}$ is the shear rate. A fundamental question of statistical mechanics, since  Boltzmann's time, is to clarify how such a chiefly dissipative transport coefficient arises from the microscopically-reversible motions of its constituents (e.g. atoms, molecules, or ions in a plasma), which is related to the well-known Loschmidt's paradox and the time's arrow problem \cite{Lebowitz,lebowitz2007timesymmetric}. 

This situation is quite different from that of the other fundamental and related property of condensed matter, the shear modulus.
The shear modulus of solids, including amorphous solids, is now well described by nonaffine response theory \cite{Scossa}, $G=G_{A}-G_{NA}$, where $G_{A}$ is the (Born) affine contribution to shear rigidity stemming from the stretching of molecular bonds, whereas $G_{NA}$ is a softening contribution due to so-called nonaffine displacements arising in disordered particle arrangements or, in crystals, due to thermal vibrations and defects. In brief, each atom or molecule is not a center of inversion symmetry, either due to intrinsic disorder of the system or due to thermal disorder. Hence, upon an applied deformation it tends to reach the new position (affine) prescribed by the macroscopic strain tenor. However, since there is no inversion symmetry, the forces from its neighbours (which are also ``en route'' towards their own affine positions) do not cancel by symmetry and thus result in a net force acting on the tagged molecule. To enforce mechanical equilibrium at all steps, this force has to be relaxed via an extra displacement, i.e. the nonaffine displacement \cite{Slonczewski,Lutsko,Lemaitre1,Scossa}. Since force times a displacement is a work, upon summing all contributions from all molecules to the free energy of deformation, the nonaffine relaxation thus contributes a negative internal work that must be subtracted from the free energy of deformation \cite{Slonczewski,Lutsko,Lemaitre1,Scossa}.

While nonaffine motions are widely accepted to play a major role in determining the shear modulus of solids (including also crystals due to instantaneous thermal random motions \cite{hong}), the question arises whether the same nonaffine response theory can also lead to a microscopic theory of the viscosity of solids and liquids.

Frameworks to compute the viscosity of liquids and solids include formalisms deduced from first principles such as the Green-Kubo formalism, which is by far the most successful and most widely employed method. 
However, these formalisms do not lead to compact closed-form expressions and typically are used for numerical calculations based on underlying molecular dynamics simulations. Furthermore, they suffer from arbitrariness in the fitting of long time tails.

In this paper, we provide a derivation of viscosity of liquids and solids from first principles within the nonaffine response formalism. This leads to a new compact formula for the viscosity of liquids and solids in terms of physical quantities that can be easily estimated either experimentally or from molecular simulations. Importantly, the derived formula provides, for the first time, a direct connection between fundamental physical quantities such as the viscosity, the speed of sound and the vibrational density of states of the system.

\section{Previous approaches}
\subsection{Green-Kubo formalism}
The most widespread recipe to compute the viscosity of a liquid in molecular dynamics is based on the Green-Kubo formalism.
The Green-Kubo relations provide a direct quantitative link between some macroscopic transport property $L$ and the time-integral of the time-correlation function of some conserved property $A$:
\begin{equation}
L= \int_0^\infty \left\langle \dot{A}(t) \dot{A}(0) \right\rangle dt.
\end{equation}
For the viscosity, the Green-Kubo relation reads as \cite{Hansen}:
\begin{equation}
    \eta = \beta V \int_{0}^{\infty}\langle \sigma_{xy}(t)\sigma_{xy}(0) \rangle dt.
\end{equation}
For a derivation of this formula see e.g. \cite{Morriss_book}.
While elegant and invaluable to compute viscosity based on MD simulations of fluids, the above formula does not lead to microscopic closed-form expressions of the kind that we discussed in the previous sections. This is because time-correlation functions such as the stress autocorrelation function, are easy to get from MD simulations but are difficult to be evaluated from a theoretical perspective, with, perhaps, the exception of Mode-Coupling theory.

The stress autocorrelation function in the above formula can be evaluated in MD simulations based on equilibrium static snapshots, i.e. without the need of deforming the simulation box. This is done by taking advantage of the virial stress formula for the stress tensor $\sigma_{xy}$ \cite{Allen_Tildesley}.

The accuracy of the Green-Kubo estimate suffers in MD simulations because a correlation function has to be integrated, and the long-time tail of the correlation is usually not sampled very well. Fitting tails on correlation functions is, indeed, a kind of an art form, and it can change the predicted transport properties quite significantly.

\subsection{The Born-Green formula}
The other fundamental microscopic expression for the liquid viscosity is the Born-Green formula \cite{Born_Green}, which establishes a fundamental link between the viscosity and the radial distribution function (rdf) $g(r)$:
\begin{equation}
\eta \sim \frac{2 \pi}{15}\left(\frac{m}{k_{B}T}\right)^{1/2}\rho^{2}\int_{0}^{\infty}g(r)\frac{dU}{dr}r^{4}dr
\end{equation}
where $m$ is the atomic mass and $U(r)$ is the interatomic pair potential. The prefactor of the formula was not determined \cite{Morioka} which makes this formula of more limited use in practical calculations. Besides the useful connection between $\eta$ and $g(r)$, this formula has the merit of recovering the Enskog formula for the viscosity of dense hard-sphere gases.

\subsection{Approaches based on activation rate theory}
Starting from the pioneering work of Eyring \cite{Eyring}, the viscosity of liquids and solids has been described using the framework of activated-rate theory of chemical reactions. Within this scheme, the viscosity is computed in terms of the hopping rate required for an atom to jump off the cage of its nearest-neighbors. The energy barrier, as clarified by Frenkel \cite{frenkel}, is represented by the energy required to re-adjust the nearest-neighbor cage, and turns out to be proportional to the high-frequency shear modulus of the system as later pointed out by Dyre in the so-called ``shoving model'' \cite{Dyre}.
Further developments based on the shoving model and on the microscopic modelling of the high-frequency shear modulus have led to closed-form expressions for the viscosity of liquids and supercooled liquids that display a double-exponential dependence on temperature, such as the KSZ model \cite{KSZ} and allow for rationalizing the fragility of chemically-different liquids based on the short-range part of the potential of mean force and the thermal expansion coefficient \cite{Lunkenheimer2023}.
While these models are invaluable for a qualitative understanding of the viscosity of liquids and supercooled liquids depending on the underlying microscopic structure and interactions, they are of limited predictive power as they contain, at best, two adjustable parameters in the comparison with experimental data.

\section{Viscoelastic linear response theory}
For an ideal elastic solid with no dissipation, the applied stress, as a function of time, and the resulting strain, as a function of time, are perfectly in phase. For a purely viscous fluid, instead, there is a 90 degree phase of the stress leading the applied strain.
This is simply because, if one takes the applied strain $\gamma(t)$ to be a sinusoidal function with frequency $\omega$, $\gamma \sim \sin \omega t$, then according to the above quoted Newton's law, $\sigma \sim \dot{\gamma}=\frac{d}{d t}\sin \omega t = \cos \omega t$. Therefore, since $\cos \theta = \sin (\theta + \frac{\pi}{2})$, it is clear that in this case the stress is not in phase with the applied strain, but is instead lagging behind by 90 degrees.
Hence, all real materials will fall within these two limits and will be characterized by their own lag phase $\delta$, which is an intrinsic material property, such that (for the case of applied strain)
\begin{equation}
\begin{split}
\gamma &= \gamma_{0}\sin \omega t,\\
\sigma &= \sigma_{0}\sin (\omega t + \delta)
\end{split}
\end{equation}
with $0 \leq \delta \leq \frac{\pi}{2}$.

According to Boltzmann's superposition principle in linear response theory, the stress at time $t$ resulting from the application of strain at previous times $t'$ is given by 
\begin{equation}
\sigma(t)=\int_{-\infty}^tG(t-t')\dot{\gamma}(t')dt
\label{A3.stress}
\end{equation}
where $G(t)$ is the time-dependent elastic modulus, also called the ``relaxation'' modulus.



Hence, with the strain rate given by $ \dot{\gamma}(t)=\omega   \gamma_0 \cos(\omega t)$ and defining $t-t'=s $, one obtains  
\begin{equation}
    \sigma(t)=\gamma_0\int_0^{\infty} \omega G(s)\cos[\omega(t-s)]ds
\end{equation} 

Using the trigonometric identity $\cos(\theta \pm \phi)=\cos(\theta)\cos(\phi) \mp \sin(\theta)\sin(\phi)$, we arrive at the expression:
\begin{equation}
\begin{split}
&\frac{\sigma(t)}{\gamma_{0}}=\left[\omega\int_0^{\infty}G(s)\sin(\omega s) ds\right] \sin(\omega t)\\
&+\left[\omega\int_{0}^{\infty}G(s)\cos(\omega s) ds\right] \cos(\omega t), \label{long}
\end{split}
\end{equation}
which thus identifies:
\begin{equation}
    \begin{split}
        G'&=\omega\int_0^{\infty}G(s)\sin(\omega s) ds\\
        G''&=\omega\int_{0}^{\infty}G(s)\cos(\omega s) ds.
    \end{split} \label{moduli}
\end{equation}
The real and imaginary part, $G'$ and $G''$, correspond to the dissipationless and to the dissipative part of the response, respectively, and they are also known as the storage (elastic) modulus and the loss (viscous) modulus.
While $G'$ is by definition in phase with the applied strain (or with the applied stress), $G''$, instead, lags behind by 90 degrees. For a generic stress wave $\sigma_{0}e^{i \omega t}$, the strain wave lagging behind would be $\gamma_{0}e^{(i \omega t +\frac{\pi}{2})}$, hence the factor $i \equiv e^{i\frac{\pi}{2}}$ in front of $G''$, leading to the complex shear modulus defined as
\begin{equation}
    G^{*}=G' + i G''.
\end{equation}
In the linear response regime, with intrinsic material properties that do not vary with time and with causality being obeyed, $G'$ and $G''$ are related to each other by the Kramers-Kronig relations.

Applying the trigonometric identity $ \sin(\theta \pm \phi)=\sin(\theta) \cos(\phi) \pm \cos(\theta) \sin(\phi)$ to $\sigma(t)=\sigma_0  \sin (\omega t + \delta) $ leads to :
\begin{equation}
\frac{\sigma(t)}{\gamma_{0}}=\frac{\sigma_0}{\gamma_0} \cos(\delta) \sin (\omega t)+ \frac{\sigma_0}{\gamma_0} \sin(\delta)  \cos (\omega  t).\label{deltas3}
\end{equation}
from which, by direct comparison with Eq. \eqref{long}, we get that $G' \propto \cos \delta$ and $G'' \propto \sin \delta$, and
\begin{equation}
    \delta = \arctan \frac{G''}{G'}.\label{arctan}
\end{equation}

Furthermore, from Eq. \eqref{long} and Eq. \eqref{moduli}, we obtain 
\begin{equation}
    \sigma(t)=G'\gamma(t) + \frac{G''}{\omega}\dot{\gamma}(t) \label{fundamental}
\end{equation}
which, by comparison with Newton's law of viscous fluids, $\sigma = \eta \dot{\gamma}$, yields the following identification:
\begin{equation}
    \eta = \frac{G''}{\omega} \label{viscosity}
\end{equation}
between the viscosity $\eta$ of the system and the loss modulus $G''$.
The first part of Eq. \eqref{fundamental} is nothing but Hooke's law and thus represents the elastic part of the response, with $G'\rightarrow G$ in the limit $G''\rightarrow 0$, where $G$ is the elastic shear modulus.
Also, for an impulse Dirac-delta applied strain $\dot{\gamma}=\gamma_{0}\delta(t)$ in Eq. \eqref{A3.stress} $G(t)=\frac{\sigma(t>0)}{\gamma_{0}}$ provides the response to a mechanical \emph{creep} test. In the rheological practice, creep measurements are often implemented by applying a constant step stress input, and measuring the strain as a function of time. When rheologists measure $G(t)$, it is an impulse strain response that is applied, and we call this material function $G(t)$ the relaxation modulus (i.e. the stress relaxes, while the strain does not change).

Until recently, most viscoelastic models have been largely phenomenological, i.e. without any connection to the microscopic physics (and chemistry) of atoms, molecules, and interactions, bonding and structuring thereof. 
The most important, historically, of such models has been the Maxwell model for viscoelastic liquids. 
This is a mathematical interpolation between the two limits of Hookean elastic solid and viscous Newtonian fluid, where an elastic Hookean element (spring) is connected in series with a dissipative element (a damper or dashpot). 
According to Maxwell'model the complex shear modulus is identified explicitly as:
\begin{equation}
    G^{*}(\omega)\equiv \frac{\sigma_0}{\gamma_0}=G'+iG''= \frac {G_{\infty}\,\tau_M^2\,\omega^2} {\tau_M^2\, \omega^2 + 1}  + i 
\frac {G_{\infty}\, \tau_M\, \omega} {\tau_M^2\, \omega^2 + 1}.\label{Maxwell_moduli} 
\end{equation}
Here we have denoted $G\equiv G_{\infty}$ to indicate that, in this case, $G$ is what is left in the response in the infinite frequency limit, $\omega \rightarrow \infty$, where, thus, the response approaches the purely elastic limit.

From a more microscopic point of view (which of course is not present in the Maxwell model, that knows nothing about the molecular structure and dynamics of matter), this is again consistent with the fact that, in the infinite frequency limit, the response of a liquid is solidlike, and this plateau in $G'$, coincides with the affine modulus $G_A$ \cite{Palyulin,Milkus2}. The fact that simple liquids respond to deformations like solids in the high frequency limit is a well known fact to anyone who dives into water from an elevated height. This fact was emphasized, in the early days of liquid state theory, by Y. Frenkel in his early monograph on liquids and amorphous solids \cite{frenkel}.
We also note that, again consistent with the phenomenology of liquids, the Maxwell model predicts $G'\rightarrow 0$ at $\omega\rightarrow 0$, i.e. the material is not rigid since it has a vanishing shear modulus at zero frequency/rate of deformation.

These facts are recovered by the Maxwell model, quite amazingly, without anything in the model which connects to the physics of real liquids and to their microscopic physics.\footnote{The above facts bear also important consequences for a broader understanding of liquids. In particular, although solids, liquids and glasses feature propagating longitudinal sound waves (though with different velocities), the dynamics of transverse (shear) waves in the three states of matter are very different. Solids exhibit propagating shear waves down to arbitrarily low momentum $k$, and their velocity is set by the shear elastic modulus $G$. Liquids, on the contrary, do not display propagating shear waves at low momenta, thus consistent with the fact that $G'=0$ at $\omega$  and $k$ going to zero. Nevertheless, propagating shear waves appear above a certain critical momentum, again consistent with the above picture provided by the Maxwell model, which has historically been called the Frenkel theory of liquids \cite{frenkel}, or more recently, the k-gap theory developed by K. Trachenko and co-workers \cite{BAGGIOLI20201}.}

\section{Microscopic nonaffine theory of viscoelasticity}
In this section, we shall extend the nonaffine elasticity theory to include dissipative effects at the microscopic level. The starting point is the writing of a suitable nonequilibrium (generalized Langevin) equation of motion for a tagged particle (atom, molecule, grain) which interacts with many other particles, the latter schematized as harmonic oscillators. This dynamical coupling, which, at the microscopic level, is mediated by the long-range part of the interatomic or intermolecular interaction forces, gives rise to a viscous-type friction term in the equation of motion, and to a stochastic thermal noise. 

\subsection{Nonequilibrium dissipative equation of motion}
Our aim is to derive a suitable equation of motion for a tagged atom (or ion, molecule, particle) in a glass in response to an applied strain. This can be done by extending the Zwanzig-Caldeira-Leggett (ZCL) \cite{Zwanzig1973,Caldeira} approach to microscopic dynamics in disordered materials. In the construction of this approach, the dynamical coupling between the tagged particle and many other particles is an effective way of describing anharmonicity, and a mapping between the ZCL model Hamiltonian and real molecular systems can be demonstrated, although this mapping is not believed to be one-to-one \cite{Rognoni}.

In the ZCL approach, the Hamiltonian of a tagged particle (mass $M$, position $Q$ and momentum $P$) coupled to all other particles (treated as harmonic oscillators with mass $M_m$, position $X_m$ and momentum $P_m$) in the material can be written as (cfr. Ref.~\cite{Zwanzig1973}):
\begin{equation}
\begin{split}
\mathcal{H}&=\frac{P^2}{2M}+U(Q)\\
&+\frac{1}{2}\sum_{m}\left[\frac{P_{m}^2}{M_{m}}+M_{m}\omega_{m}^2\left(X_{m}
-\frac{\gamma_{m}Q}{M_{m}\omega_{m}^2}\right)^{2}\right].\label{Hamiltonian}
\end{split}
\end{equation}
This formalism can be extended to include the presence of externally applied field. The term describing the dynamical coupling between the tagged particle and the $m$th-oscillator is defined as $\gamma_m$.  
Introducing the mass-scaled tagged-particle displacement $s=Q\sqrt{M}$, the resulting generalized Langevin equation of motion for the displacement of the tagged particle becomes (cfr. \cite{Zwanzig1973} for the full derivation):
\begin{equation}
\ddot{s}=-U'(s)-\int_{-\infty}^t \nu(t-t')\frac{ds}{dt'}dt' + F_{p}(t),
\label{2.4gle1}
\end{equation}
where $F_{p}(t)$ is the thermal stochastic noise with zero average, $U$ is a local interaction potential (e.g. with the nearest neighbours), and $\nu$ is the friction resulting from many long-range interactions with all other particles in the system, imposed by the dynamical bi-linear coupling.
For dynamical response to an oscillatory strain, one can average the dynamical equation over many cycles, which amounts to a time-average \cite{PhysRevB.95.054203,Cui_GLE}. Since the noise $F_{p}$ is defined to have zero-mean \cite{Zwanzig1973}, an average over many cycles leaves $\langle F_{p} \rangle=0$ in the above equation.\footnote{According to Ref. \cite{PhysRevB.95.054203}, when the system is non-ergodic below $T_{g}$, nothing guarantees this is true a priori, but there is evidence that this approximation might be reasonable in the linear regime where the response converges to a reproducible noise-free average stress.}

\subsection{Derivation of microscopic viscoelastic moduli}
We rewrite the Eq. \eqref{2.4gle1} for a tagged atom in $d$-dimensions, which moves with an affine velocity prescribed by the strain-rate tensor
$\dot{\mathbf{F}}$ (where the dot indicates a time derivative while the circle indicates quantities measured in the undeformed rest frame):
\begin{equation}
\ddot{\mathbf{r}}_i^\mu=\mathbf{f}_i^\mu-\int_{-\infty}^t\nu(t-t')\left(\dot{\mathring{\mathbf{r}}}_i^\mu - \mathbf{u}^\mu\right) dt'
\end{equation}
where $\mathbf{f}_i^\mu=-\partial{U}/\partial{\mathbf{r}}_i^\mu$ generalises the $-U'(s)$ to the tagged atom.
Furthermore, we used the Galilean transformations to express the particle velocity in the moving frame: $\dot{\mathbf{r}}_{i}=\dot{\mathring{\mathbf{r}}}_i - \mathbf{u}$ where $\mathbf{u}=\dot{\mathbf{F}}\mathbf{\mathring{r}}_{i}$ represents the local velocity of the moving frame. 
Since we are going to work at constant deformation rates, hence at constant velocity of the moving frame, the Galilean transformation valid for inertial frames is appropriate.
This notation is consistent with the use of the circle on the particle position variables to signify that they are measured with respect to the reference rest frame. In terms of the original rest frame $\lbrace\mathbf{\mathring{r}}_i\rbrace$, the equation of motion can be written, for the particle position averaged over several oscillations, as
\begin{equation}
\mathbf{F}\,\mathbf{\ddot{\mathring{r}}}_i  =\mathbf{f}_i-
\int_{-\infty}^{t}\nu(t-t')\cdot\frac{d\mathring{\mathbf{r}}_i}{dt'}dt',
\label{2.4restframe}
\end{equation}
where $\langle F_{p} \rangle=0$ was dropped for the reasons explained above.\footnote{Terms $\ddot{\mathbf{F}}\mathbf{\mathring{r}}_i$ and $\int_{-\infty}^{t}\nu(t-t')\dot{\mathbf{F}}\cdot\mathring{\mathbf{r}}_idt'$ are not allowed to enter the equation of motion because
they depend on the position of the particle, and therefore have to vanish for a system with translational invariance, as noted already in \cite{Andersen} and in \cite{Ray1984}.}

We work in the linear regime of small strain $\parallel\mathbf{F}-\mathbf{1}\parallel \ll 1$ by making a perturbative expansion in the small displacement $\{\mathbf{s}_i(t)=\mathbf{\mathring{r}}_i(t)-\mathbf{\mathring{r}}_i\}$ around a known rest frame $\mathbf{\mathring{r}}_i$.
That is, we take $\mathbf{F}=\mathbf{1}+\delta\mathbf{F}+...$ where $\delta\mathbf{F}\approx\mathbf{F}-\mathbf{1}$ is the small parameter. Replacing this back into Eq. \eqref{2.4restframe} gives
\begin{align}
&(\mathbf{1}+\delta\mathbf{F}+...)\frac{d^2\mathbf{s}_i}{dt^2}=\delta\mathbf{f}_i+\nonumber\\
&-(\mathbf{1}+\delta\mathbf{F}+...)\int_{-\infty}^{t}\nu(t-t')\cdot\frac{d\mathbf{s}_i}{dt'}dt'.
\label{2.4expansion}
\end{align}
For the term $\delta\mathbf{f}_i$, imposing mechanical equilibrium again, which is $\mathbf{f}_{i}=0$,  
implies:
\begin{equation}
    \delta \mathbf{f}_{i} = \frac{\partial \mathbf{f}_{i}}{\partial \mathbf{\mathring{r}}_j }\delta \mathbf{\mathring{r}}_j + \frac{\partial \mathbf{f}_{i}}{\partial \mathbf{\eta}}:\delta \mathbf{\eta}
\end{equation}
where in the first term we recognise
\begin{equation}
    \frac{\partial \mathbf{f}_{i}}{\partial \mathbf{\mathring{r}}_j }\delta \mathbf{\mathring{r}}_j =-\mathbf{H}_{ij}\mathbf{s}_{j}
\end{equation}
while for the second term we have:
\begin{equation}
    \mathbf{\Xi}_{i,\kappa\chi}=\frac{\partial \mathbf{f}_{i}}{\partial \mathbf{\eta}_{\kappa\chi}}
    \label{aff_force}
\end{equation}
and the limit $\mathbf{\eta}_{\kappa\chi} \rightarrow 0$ is implied. Here $\eta_{\kappa\chi}$ are the components of the Cauchy-Green strain tensor defined as $\mathbf{\eta} =\frac{1}{2}\left(\mathbf{F}^{T}\mathbf{F}-\mathbf{1} \right)$. This is a second-rank tensor, and should not be confused with the fluid viscosity $\eta$ (a scalar).
Using standard lattice dynamics, the affine force Eq. \eqref{aff_force} can be evaluated for a shear deformation $\kappa\chi=xy$ as \cite{Lemaitre1}
\begin{equation}
\mathbf{\Xi}_{i,xy} = -\sum_{j}\left( \kappa_{ij}r_{ij} - t_{ij}\right) n_{ij}^{x} n_{ij}^{y} \hat{\mathbf{n}}_{ij},
\label{xi_lattice}
\end{equation}
where $\kappa_{ij}$ is the bond spring constant between particles $i$ and $j$, as standard in lattice dynamics, $t_{ij}$ is the bond tension (first derivative of the interaction energy evaluated at the interparticle distance), and $r_{ij}$ is the modulus of the interparticle distance. Here the centrosymmetry or lack thereof is contained in the sum: 
$\sum_{j}...n_{ij}^{x} n_{ij}^{y} \hat{\mathbf{n}}_{ij}$
where $\hat{\mathbf{n}}_{ij}$ is the unit vector from atom $i$ to at any other atom $j$, while $n_{ij}^{x}$ and $n_{ij}^y$ are its $x$ and $y$ components, respectively.
It is clear that this sum, and hence the force $\mathbf{\Xi}_{i,xy}$, will be zero for a centrosymmetric arrangement of the atoms $j$ around the atom $i$, and hence there will be no nonaffine motions. In a non-centrosymmetric lattice, as well as in a liquid or in a glass, the above factor will rarely be zero, and therefore the force  $\mathbf{\Xi}_{i,xy}$ will almost always be there. For more complex lattices, e.g. non-Bravais lattices, the force $\mathbf{\Xi}_{i,xy}$ has to be evaluated precisely based on the atomic positions in the lattice, e.g. based on the known static structure or atomistic simulation snapshot.

With these identifications, we can write Eq. (\ref{2.4expansion}), to first order:
\begin{equation}
\frac{d^2\mathbf{s}_i}{dt^2}+\int_{-\infty}^{t}\nu(t-t')
\frac{d\mathbf{s}_i}{dt'}dt'+\mathbf{H}_{ij}\mathbf{s}_{j}=\mathbf{\Xi}_{i,\kappa\chi}\eta_{\kappa\chi}.
\label{2.4gle2}
\end{equation}

The above equation can be solved by Fourier transformation followed by a normal mode decomposition, as we shall see next. If we specialise on time-dependent uniaxial strain along the $x$ direction, $\eta_{xx}(t)$, then the vector $\mathbf{\Xi}_{i,xx}\eta_{xx}(t)$ represents the force acting on particle $i$ due to the motion of its nearest-neighbors which are moving towards their respective affine positions (see e.g.~\cite{Lemaitre1} for a more detailed discussion). Hence, all terms in the above Eq. \eqref{2.4gle2} are vectors in $\mathbb{R}^{3}$ and the equation is in manifestly covariant form.
In metals, this ``drag force'' also includes electronic effects taken into account semi-empirically e.g. via the embedded-atom model (EAM) \cite{Finnis}.

To make it convenient for further manipulation, we extend all matrices and vectors to $Nd \times Nd$ and $Nd$-dimensional, respectively, and we will select $d=3$.
After applying Fourier transformation to Eq. \eqref{2.4gle2}, we obtain
\begin{equation}
-\omega^2\,\tilde{\mathbf{s}}+i\tilde{\nu}(\omega)\omega\,\tilde{\mathbf{s}}
+\mathbf{H}\,\tilde{\mathbf{s}}
=\mathbf{\Xi}_{\kappa\chi}\tilde{\eta}_{\kappa\chi},
\end{equation}
where $\tilde{\nu}(\omega)$ is the Fourier transform of $\nu(t)$ etc (we use the tilde consistently throughout to denote Fourier-transformed quantities). In the above equation, all the terms are are now vectors in $\mathbb{R}^{3N}$ space.
Next, we apply normal mode decomposition in $\mathbb{R}^{3N}$ using the $3N$-dimensional eigenvectors of the Hessian as the basis set for the decomposition. This is equivalent to diagonalise the Hessian matrix $\mathbf{H}$. 
Proceeding in the same way as in \cite{Cui_viscoelastic}, we have that the $m$-th mode of displacement can be written as:
\begin{equation}
-\omega^2\hat{\tilde{s}}_m(\omega)+i\tilde{\nu}(\omega)\omega\,\hat{\tilde{s}}_m(\omega)
+\omega_m^2\hat{\tilde{s}}_m(\omega)
=\hat{\Xi}_{m,\kappa\chi}(\omega)\tilde{\eta}_{\kappa\chi}.
\label{2.4kerneldecom}
\end{equation}
It was shown \cite{Lemaitre1}, by means of MD simulations for a Lennard-Jones system, that $\hat{\Xi}_{m,\kappa\chi}=\mathbf{v}_m\cdot\mathbf{\Xi}_{\kappa\chi}$ is self-averaging (even in the glassy state), and one might therefore introduce the smooth correlator function on eigenfrequency shells
\begin{equation}
\Gamma_{\mu\nu\kappa\chi}(\omega)=\langle\hat{\Xi}_{m,\mu\nu}\hat{\Xi}_{m,\kappa\chi}\rangle_{\omega_m\in\{\omega,\omega+d\omega\}} \label{Gamma}
\end{equation}
on frequency shells.
Following the general procedure of ~\cite{Lemaitre1} to find the oscillatory stress for a dynamic nonaffine deformation, the stress is obtained to first order in strain amplitude as a function of $\omega$ (note that the summation convention is active for repeated indices):
\begin{align}
\tilde{\sigma}_{\mu\nu}(\omega)&=C^A_{\mu\nu\kappa\chi}\tilde{\eta}_{\kappa\chi}(\omega)-\frac{1}{\mathring{V}}\sum_{m}\hat{\Xi}_{m,\mu\nu}\hat{\tilde{s}}_m(\omega) \notag\\
&=C^A_{\mu\nu\kappa\chi}\tilde{\eta}_{\kappa\chi}(\omega)-\frac{1}{\mathring{V}}\sum_{m}\frac{\hat{\Xi}_{m,\mu\nu}\hat{\Xi}_{m,\kappa\chi}}{\omega_{m}^2-\omega^2
+i\tilde{\nu}(\omega)\omega}\tilde{\eta}_{\kappa\chi}(\omega)\notag\\
&\equiv C_{\mu\nu\kappa\chi}(\omega)\tilde{\eta}_{\kappa\chi}(\omega).\label{generalized}
\end{align}

In the thermodynamic limit and assuming a continuous vibrational spectrum, we can replace the discrete sum over $3N$ degrees of freedom with an integral over vibrational frequencies up to the Debye (cut-off) frequency $\omega_{D}$. In this case, we need to replace the discrete sum over the $3N$ degrees of freedom (eigenmodes) with an integral, $\sum_{m=1}^{3N}...\rightarrow \int_{0}^{\omega_D}g(\omega_p)...d\omega_p$, where $g(\omega_p)$ is the vibrational density of states (VDOS).

Then, the complex elastic constants tensor can be obtained as:
\begin{equation}
C_{\mu\nu\kappa\chi}(\omega)=C^A_{\mu\nu\kappa\chi}-3\rho\int_0^{\omega_{D}}\frac{g(\omega_p)\Gamma_{\mu\nu\kappa\chi}(\omega_p)}{\omega_p^2-\omega^2+i\tilde{\nu}(\omega)\omega}d\omega_p
\label{2.4modulus}
\end{equation}
where $\rho=N/\mathring{V}$ denotes the number density of the solid in the initial state. One should note that, if the VDOS is normalized to $3N$, as often done in the literature \cite{Kostya_book}, then $\rho=1/\mathring{V}$ and there is no factor $3$ in the above equation.
This is a crucial result obtained in \cite{Cui_viscoelastic}, which differs from a previous result obtained in \cite{Lemaitre1} because the friction is non-Markovian, hence frequency or time/history-dependent, whereas in \cite{Lemaitre1} it is just a constant, corresponding to Markovian dynamics. This turns out to be a fundamental difference, because  experimental data of real materials cannot be described by a constant friction coefficient \cite{Cui_viscoelastic}. 
In the above we used mass-rescaled variables throughout, so that the atomic/molecular mass is not present explicitly in the final expression.
If one, instead, uses non-rescaled variables and specializing to shear deformations, $\mu\nu\kappa\chi=xyxy$, we obtain the following expressions for the complex shear modulus $G^{*}$:
\begin{equation}
G^{*}(\omega)=G_A-3\rho\int_0^{\omega_{D}}\frac{g(\omega_p)\Gamma_{xyxy}(\omega_p)}{m\omega_p^2-m\omega^2+i\tilde{\nu}(\omega)\omega}d\omega_p.
\label{2.4shear_modulus}
\end{equation}
In the above expression the frequencies are now in physical units of Hertz. The first term on the r.h.s. is the affine shear modulus $G_A$, which is independent of $\omega$.
The low-frequency behaviour of $\Gamma(\omega_p)$ can be estimated analytically using the following result:
\begin{equation}
\langle\hat{\Xi}_{p,\mu\nu}\hat{\Xi}_{p,\kappa\chi}\rangle = d\kappa
R_0^2\, \lambda_{p}\sum_{\alpha}B_{\alpha,\mu\nu\kappa\chi},\label{correlator}
\end{equation}
derived originally in Ref. \cite{Scossa}, which gives $\langle \hat{\Xi}_{p,xy}^{2}\rangle \propto \lambda_{p}$, thus implying (from its definition above): 
\begin{equation}
\Gamma(\omega_p) \propto \omega_{p}^{2}.\nonumber
\end{equation}
This analytical estimate appears to work reasonably well in the low-eigenfrequency part of the $\Gamma(\omega_p)$ spectrum of amorphous solids \cite{Palyulin,Lemaitre1,Milkus2}. 
Furthermore, $\sum_{\alpha}B_{\alpha,\mu\nu\kappa\chi}$ is a geometric coefficient that depends only on the geometry of macroscopic deformation and its values can be found tabulated in Ref. \cite{Scossa}.

By separating real and imaginary part of the above expression, we then get to the storage and loss moduli as \cite{Milkus2}:
\begin{equation}
    \begin{split}
    G'(\omega)&=G_A - 3\rho\int_{0}^{\omega_{D}}\frac{m\,g(\omega_p)\,\Gamma(\omega_p)\,(\omega_{p}^{2}-\omega^{2})}{m^{2}(\omega_{p}^{2}-\omega^{2})^{2}+\tilde{\nu}(\omega)^{2}\omega^{2}}d\omega_p\\
    G''(\omega)&=3\rho\int_{0}^{\omega_{D}}\frac{g(\omega_p)\,\Gamma(\omega_p)\,\tilde{\nu}(\omega)\,\omega}{m^{2}(\omega_{p}^{2}-\omega^{2})^{2}+\tilde{\nu}(\omega)^{2}\omega^{2}}d\omega_p.\label{visco-moduli}
    \end{split}
\end{equation}
It is easy to check that the storage modulus $G'(\omega)$ reduces to the Born modulus $G_A \equiv G_\infty$ in the infinite-frequency limit, $\omega \rightarrow 0$.
From the point of view of practical computation, the VDOS $g(\omega_p)$ can be obtained numerically via direct diagonalization of the Hessian matrix $\mathbf{H}_{ij}$, since its eigenvalues are related to the eigenfrequencies via $\lambda_p=m\omega_{p}^{2}$. Similarly, the affine-force correlator $\Gamma(\omega_p)$ can also be computed from its definition by knowing the positions of all the particles, their interactions and forces (so that the affine force fields $\mathbf{\Xi}$ can be computed) as well as the eigenvectors of the Hessian $\mathbf{v}_p$.
Calculating eigenvalues and eigenvectors of the Hessian matrix is a computationally demanding task, especially in terms of RAM. Direct diagonalization (DD) of the Hessian is feasible on standard computers only up to $N \sim 10^4$ particles/atoms. 
As shown in Ref. \cite{Ivan}, the RAM usage for direct diagonalization of the Hessian scales with the number of atoms as $\sim N^{2}$, i.e. very unfavourably. The computational time scales also very unfavourably, as $\sim N^{2.71}$.
As a way to obviate this problem, a computational protocol based on the Kernel Polynomial Method (KPM) has been developed in \cite{Ivan}, which is based on approximating eigenvector-based quantities of the Hessian with Chebyshev polynomials. With this methodology, it is possible to have a much more favourable scaling, i.e. linear in $N$, for both the RAM and the computational time. This makes it possible to compute $g(\omega_p)$ and $\Gamma(\omega_p)$ for much larger systems, i.e. $N > 10^5$, which would otherwise be impossible with direct diagonalization.  

Importantly, for simple liquids at thermodynamic equilibrium, it has been demonstrated in Ref. \cite{wittmer}, by using nonaffine response theory combined with equilibrium statistical mechanics, that the affine shear modulus $G_{A}$ is identical, with opposite sign, to the nonaffine part, thus resulting in $G'(\omega=0)=0$ for liquids at equilibrium. This is an important check for the correctness and generality of the nonaffine deformation theory.

\section{Viscosity from microscopic nonaffine response theory}
We recall that the viscosity can be obtained from the loss viscoelastic modulus $G''$ using nonaffine response theory as (cfr. Eq. \eqref{viscosity}):
\begin{equation}
    \eta = \frac{G''}{\omega}.
\end{equation}

The nonaffine response theory developed from first principles in section IV provides the following form for the loss modulus $G''$ (cfr. Eq. \eqref{visco-moduli}):
\begin{equation}
 G''(\omega)=3\rho\int_{0}^{\omega_{D}}\frac{g(\omega_p)\,\Gamma(\omega_p)\,\tilde{\nu}(\omega)\,\omega}{m^{2}(\omega_{p}^{2}-\omega^{2})^{2}+\tilde{\nu}(\omega)^{2}\omega^{2}}d\omega_p. \label{loss_repeated}
 \end{equation}
For shear deformation ($\kappa\chi=xy$), 
Eq. \eqref{generalized} becomes:
\begin{equation}
    \sigma_{xy}(\omega)=G^{*}(\omega)\gamma(\omega).
\end{equation}
Since $G^{*}=G' + iG''$ and there is factor $\omega$ in the above expression for $G''$ in Eq. \eqref{loss_repeated}, the theory correctly recovers, for the dissipative part of the stress, $\sigma_{xy}'$, the Newton law of viscous flow (cfr. Newton's law of viscous liquids $\sigma' = \eta \dot{\gamma}$):
\begin{equation}
    \sigma_{xy}' = \eta \, i \omega \gamma = \eta \dot{\gamma}
\end{equation}
with, indeed, a zero-frequency shear viscosity given by:
\begin{equation}
    \eta = 3\rho\tilde{\nu}(0)\int_{0}^{\omega_{D}}\frac{g(\omega_p)\,\Gamma(\omega_p)}{m^{2}\omega_{p}^{4}}d\omega_p. \label{viscosity_nonaffine}
\end{equation}
This formula provides a direct, and unprecedented, connection between the viscosity $\eta$ and the vibrational density of states (VDOS) $g(\omega_p)$. This is the most important result of this work. The VDOS is an experimentally measurable quantity that can be obtained from inelastic neutron scattering \cite{Dehong} or from Raman scattering \cite{Shuker}.
Furthermore, this formula shows that the non-kinetic part of the viscosity goes to zero whenever the memory function or spectral function goes to zero at zero frequency, i.e. when $\tilde{\nu}(\omega \rightarrow 0)=0$. This is physically meaningful because a nearly dissipationless fluid with the non-kinetic part of the viscosity equal to zero can only arise when the correlations decay completely to zero in the long-time limit.

For a solid, which follows the Debye law Eq. \eqref{Debyedens}: 
\begin{equation}
g(\omega_{p}) = \frac{\omega_{p}^2\,V}{2\,\pi^2}\,\left(\frac{2}{v_T^3}\,+\,\frac{1}{v_L^3}\right) \label{Debyedens}
\end{equation}
and for which $\Gamma(\omega_{p})\sim \omega_{p}^{2}$ as derived in \cite{Scossa}, we obtain the simple and compact relation (valid up to some undetermined numerical prefactor):
\begin{equation}
    \eta \approx \rho\tilde{\nu}(0) \frac{V}{2\,\pi^2}\,\left(\frac{2}{v_T^3}\,+\,\frac{1}{v_L^3}\right)\frac{\omega_{D}}{m^{2}}.\label{simplified}
\end{equation}
Although highly idealized, this simple microscopic relation provides a direct proportionality law between the viscosity $\eta$ and important physical quantities: the zero-frequency limit of the Fourier-transformed friction kernel, $\tilde{\nu}(0)$,\footnote{which in turn arises from anharmonic couplings between each particle and the other particles via the Caldeira-Leggett or ZCL Hamiltonian} the Debye frequency and the longitudinal and transverse speeds of sound. 
This formula Eq. \eqref{simplified} is also correct from a dimensional point of view: upon recalling Eq. \eqref{correlator}, the dimensionality of $\Gamma(\omega_p)$ is given by $\Gamma(\omega_p) \sim \kappa R_{0}^{2} \lambda_{p} \sim m\kappa R_{0}^{2} \omega_{p}^{2} $, where dimensionless factors have been omitted. We thus get:
\begin{equation}
    \eta = \kappa R_{0}^{2} \rho\,\tilde{\nu}(0) \frac{V}{2\,\pi^2}\,\left(\frac{2}{v_T^3}\,+\,\frac{1}{v_L^3}\right)\frac{\omega_{D}}{m},\label{simplified2}
\end{equation}
where the spring constant $\kappa$ has dimensions of force per unit length $\kappa \sim [F/L]$, while $R_0 \sim [L]$ is an atomic-scale length-scale, and the memory kernel at zero frequency has dimension of a Langevin-type friction, i.e. mass times inverse time, $\tilde{\nu}(0) \sim [m \cdot \frac{1}{s}]$. Using these facts, the dimensional analysis gives $\eta$ correctly as as a viscosity, i.e. force per unit area times inverse time, $\eta \sim [\frac{F}{L^2} \cdot s]$, in SI units $[Pa \cdot s]$.
Furthermore, the above simplified formula Eq. \eqref{simplified}, exhibits the same linear dependence of viscosity on the Debye frequency as recently proposed in Ref. \cite{Trachenko2020}.

More generally, realistic predictions can be made using the full Eq. \eqref{viscosity_nonaffine} with the VDOS $g(\omega_{p})$ computed for a realistic system by means of MD simulations or from experimental data. Also $\Gamma(\omega_{p})$ has to be evaluated numerically based on the eigenvectors of the Hessian matrix, although for disordered solids the law $\Gamma(\omega_{p})\sim \omega_{p}^{2}$ might be a reasonable approximation to deduce analytical correlations \cite{Palyulin,Milkus2}.

For liquids, the VDOS $g(\omega_{p})$ can be computed via MD simulations, and only recently it has been measured experimentally with inelastic neutron scattering in Ref. \cite{Dehong}. The main complication is that the VDOS of 
liquids contains a large population of Instantaneous Normal Modes (INMs), cfr.  \cite{Keyes_INMs,Stratt_INMs,Palyulin,Douglas_INMs}, and these need be taken into account in evaluating the integral in Eq. \eqref{viscosity_nonaffine} following the method of Refs. \cite{Palyulin,Ivan}.
Recently, an analytical theory of the VDOS of liquids accounting for INMs has been developed in Ref. \cite{Baggioli_PNAS} and experimentally confirmed in Ref. \cite{Dehong}.

Finally, $\tilde{\nu}(0)$ represents the zero-frequency limit of the one-sided Fourier transform, $\tilde{\nu}(\omega)$ (also known as the ``spectral density'' \cite{Weiss}) of the friction memory kernel $\nu(t)$. There are various ways of estimating $\nu(t)$. Starting from the particle-bath Hamiltonian cfr. Eqs. \eqref{Hamiltonian}, upon integrating the Euler-Lagrange equations for the coupled dynamics of the tagged particle and the heat-bath oscillators, Zwanzig obtained the identification \cite{Zwanzig1973}:
\begin{equation}
    \nu(t)= \sum_{m}\frac{\gamma_{m}^{2}}{m\omega_{m}^{2}} \cos(\omega_{m}t) \label{friction_series}
\end{equation}
where the $\gamma_{m}$ is the coupling coefficient in Eq. \eqref{Hamiltonian} between the tagged particle and the $m$-th bath oscillator.
As usual, one can then move from the discrete set of oscillators frequencies to a continuous integral over the VDOS, leading to Eq. \eqref{3.2frictioncontinuous}:
\begin{equation}
    \nu(t) = \int_{0}^{\infty}g(\omega_{p})\frac{\gamma(\omega_{p})^{2}}{m\omega_{p}^{2}} \cos(\omega_{p}t)d\omega_{p} \label{3.2frictioncontinuous}
\end{equation}
where $\gamma(\omega_p)$ represents the continuous limit (spectrum) of the discrete set of dynamic coupling constants $\{\gamma_{p}\}$ in the ZCL Hamiltonian Eq. \eqref{Hamiltonian}. 
According to this equation, then, we would have a double dependence of the viscosity $\eta$ on the VDOS through Eq. \eqref{viscosity_nonaffine}.

The memory function or friction kernel $\nu(t)$ can be evaluated on the basis of MD simulations for which different methods are available. 
The most immediate way is to apply the fluctuation dissipation theorem (FDT) associated with the governing equation of motion, i.e. the generalized Langevin equation Eq. \eqref{2.4gle2}. The FDT is derived in \cite{Zwanzig1973} and follows as:
\begin{equation}
    \langle F_P(t)F_P(t')\rangle =k_B T\nu(t-t')  \label{FDT}
\end{equation}
where $\langle F_P(t)F_P(t')\rangle$ is the time autocorrelation function of the stochastic force $F_{p}(t)$.\footnote{The averaged time autocorrelation function of a physical variable $A$ is defined as $\langle A(t)A(0)\rangle = \frac{1}{\tau}\int_{0}^{\tau}A(t+t')A(t')dt'$. For equilibrium ergodic systems the time averaging can be replaced with an ensemble average.}
Therefore, by measuring  $\langle F_P(t)F_P(t')\rangle$, the above Eq. \eqref{FDT} allows one to determine $\nu(t)$ from MD simulations.

Since, in practice, it is easier to measure velocity autocorrelation functions, $\langle v(t)v(0) \rangle$, the following identity (obtained through integration by parts) becomes very useful \cite{Schmidt}:
\begin{equation}
    \langle F_P(t)F_P(t')\rangle = -M^{2}\frac{\partial^{2}}{\partial t^{2}}\langle v(t)v(t') \rangle
\end{equation}
where, obviously, $F_{p}$ and $v$ are, in real situations, 3D vectors. Numerical examples of this kind of reconstruction technique for the memory kernel are discussed in Ref. \cite{Schmidt}.

Alternatively, one can deduce an equation involving the momentum autocorrelation function (MAF) $C_{pp}(t)=\langle p(t) p(0) \rangle$ and the system force-momentum correlation function (MFC) $C_{pF}(t)=\langle F(t) p(0) \rangle$, where $F(t)$ denotes the conservative force acting on the tagged particle, i.e. $F \equiv -U'(s(t))$, with reference to Eq. \eqref{2.4gle2} or $\mathbf{H}_{ij}(t)\mathbf{s}_{j}(t)$ with reference to Eq. \eqref{2.4gle2}. The equation reads as \cite{Kuehn_2016}
\begin{equation}
    \dot{C}_{pp}(t)=C_{pF}(t) - \int_{0}^{t}\nu(t-t')C_{pp}(t)dt', \label{Kuehn1}
\end{equation}
from which, upon Fourier transformation, one gets:\\

\begin{equation}
\tilde{\nu}(\omega)=\frac{1+\tilde{C}_{pF}(\omega)}{\tilde{C}_{pp}(\omega)} -i\omega. \label{Kuehn}
\end{equation}
\\

Here both MFC and MAF are normalized to $C_{pp}(0)$. Computing $\tilde{\nu}(\omega)$ from molecular simulations using Eq. \eqref{Kuehn} is straightforward since all the quantities involved are readily available from MD simulations. The above Eqs. \eqref{Kuehn1}-\eqref{Kuehn} have been quoted from Ref. \cite{Kuehn_2016} with no modification, hence the units of $\nu$ and $\tilde{\nu}$ miss a mass factor compared to the units used in this paper throughout.

\section{Conclusion}
In conclusion, we have presented a first-principles derivation of the viscosity coefficient from a microscopically-reversible Caldeira-Leggett Hamiltonian, within the framework of nonaffine response theory. 
The final result is a fundamental new expression for the viscosity of liquids and solids, given by Eq. \eqref{viscosity_nonaffine}. This fundamental relation shows that the viscosity is directly proportional to the zero-frequency limit of the memory kernel for the underlying atomic-scale motion, which can be evaluated in simulations by computing time autocorrelation functions of the velocity. Compared to the Green-Kubo formula, this is certainly a substantial improvement since it avoids having to compute time autocorrelation functions of the stress (which involves fitting of long-time tails with unavoidable arbitrariness). Furthermore, the viscosity is proportional to an integral over the vibrational density of states (VDOS) of the system, which is an experimentally accessible quantity. To the best of our knowledge, this is the first time that the viscosity is quantitatively linked to the vibrational spectrum of the system. A further simplified formula for the case of solids, Eq. \eqref{simplified} provides a new link between fundamental physical quantities of condensed matter, such as the viscosity, the zero-frequency value of the spectral function or memory kernel, the Debye frequency, and the (longitudinal and transverse) speed of sound.

In future work, the above results can be tested using molecular simulations. The presented framework can also be extended to relativistic fluids, by leveraging on the relativistic extension of the generalized Langevin equation \cite{Petrosyan_2022} with important applications to the viscosity of heavy-ion beams \cite{Gardim} and quark-gluon plasmas \cite{quark}. Similarly, using the quantum version of the Caldeira-Leggett Hamiltonian \cite{Caldeira,Weiss}, the same approach can be used to compute the viscosity of quantum fluids in future work.
It can also be used in the context of granular fluids, in particular granular jets with non-trivial liquidlike properties \cite{Nagel2,Nagel1}.
Finally, the present approach can be highly complementary to ``rheological universal differential equations'' recently proposed in Ref. \cite{lennon2022scientific}.

\subsection*{Acknowledgments} 
The author is indebted to Kostya Trachenko, Gareth H. McKinley and Eugene M. Terentjev for critical reading of the preliminary manuscript and for valuable input.
I am also grateful to Prof. Vladimir Lisy and to Dr. Vinay Vaibhav for help in correcting a few typos that were present in the published version and have been corrected in this revised ArXiv version.
The author gratefully acknowledges funding from the European Union through Horizon Europe ERC Grant number: 101043968 ``Multimech'', and from US Army Research Office through contract nr.   W911NF-22-2-0256. 
\bibliographystyle{apsrev4-1}

\bibliography{refs}

\end{document}